\documentclass[superscriptaddress,prb,preprint,showpacs,showkeys,nobibnotes]{revtex4}

\usepackage{times}
\usepackage{color}
\usepackage{graphicx}
\usepackage{textcomp}
\usepackage{amsmath,amsbsy,amsfonts,mathrsfs}

\newcommand{\be}{\begin{equation}}
\newcommand{\ee}{\end{equation}}
\newcommand{\bea}{\begin{eqnarray}}
\newcommand{\eea}{\end{eqnarray}}
\newcommand{\bg}{\begin{figure}}
\newcommand{\eg}{\end{figure}}
\newcommand{\bi}{\begin{itemize}}
\newcommand{\ei}{\end{itemize}}
\newcommand{\refeq}[1]{eq. (\ref{eq:#1})}
\newcommand{\reffig}[1]{Fig. \ref{fig:#1}}

\begin{document}

\title[Mdet]{Photo-induced reactions from efficient molecular dynamics with electronic transitions using the FIREBALL local-orbital density functional theory formalism}
\author{\bf Vladm\'ir Zoba\v{c}}
\affiliation{Institute of Physic, Academy of Sciences of the Czech Republic, Cukrovarnick\'{a} 10, CZ-16200, Prague, Czech Republic}
\author{\bf James~P.~Lewis}
\email{james.lewis@mail.wvu.edu}
\affiliation{Institute of Physic, Academy of Sciences of the Czech Republic, Cukrovarnick\'{a} 10, CZ-16200, Prague, Czech Republic}
\affiliation{Department of Physics and Astronomy, West Virginia University, Morgantown WV 26506, USA}
\author{\bf Enrique Abad}
\affiliation{Computational Biochemistry Group, Institute of Theoretical Chemistry, University of Stuttgart, Pfaffenwaldring 55 70569 Stuttgart, Germany}
\author{\bf Jes\'us I. Mendieta-Moreno}
\affiliation{Departamento de F\1sica Te\'orica de la Materia Condensada and Condensed Matter Physics Center (IFIMAC),
Universidad Aut\'onoma de Madrid, ES-28049 Madrid, Spain}
\author{\bf Prokop Hapala}
\affiliation{Institute of Physic, Academy of Sciences of the Czech Republic, Cukrovarnick\'{a} 10, CZ-16200, Prague, Czech Republic}
\author{\bf Pavel Jel\'inek}
\email{jelinekp@fzu.cz}
\affiliation{Institute of Physic, Academy of Sciences of the Czech Republic, Cukrovarnick\'{a} 10, CZ-16200, Prague, Czech Republic}
\affiliation{Graduate School of Engineering, Osaka University 2-1, Yamada-Oka, Suita, Osaka 565-0871, Japan}
\author{\bf Jos\'e Ortega}
\email{jose.ortega@uam.es}
\affiliation{Departamento de F\1sica Te\'orica de la Materia Condensada and Condensed Matter Physics Center (IFIMAC),
Universidad Aut\'onoma de Madrid, ES-28049 Madrid, Spain}
\date{\today}


\begin{abstract}
The computational simulation of photo-induced processes in large molecular systems is a very challenging problem. 
Here, we present a detailed description of our implementation of a molecular dynamics with electronic transitions algorithm within the local-orbital density functional theory code {\sc Fireball}, suitable for the computational study of these problems.
Our methodology enables simulating photo-induced reaction mechanisms over hundreds of trajectories; therefore, large statistically significant ensembles can be calculated to accurately represent a reaction profile.
As an example of the application of this approach, we report results on the [2+2] cycloaddition of ethylene
with maleic anhydride and on the [2+2] photo-induced polymerization reaction of two C$_{60}$ molecules.
We identify different deactivation channels of the initial electron excitation, depending on the time of the electronic transition from LUMO to HOMO, and the character of the HOMO after the transition.
\end{abstract}



\maketitle
\section{Introduction}
One of the greatest challenges in simulation methodology is to accurately and efficiently interpret photo-induced reaction mechanisms which are critically important in many chemical processes, such as vision, photodamage, UV absorption, photosynthesis, photography, etc. In Born-Oppenheimer (or adiabatic) molecular dynamics - the nuclei follow classical trajectories defined on a single potential energy surface (PES); this PES corresponds to the ground state electronic energy of the system, for frozen atomic configurations. To accurately simulate photo-induced reactions, the excited potential energy surface must be appropriately accessed beyond the Born-Oppenheimer approximation.  Additionally, the ability of the electron to transition between the ground PES and the excited PES at the point of the conical intersection between the two PESs is very important in accurately portraying the transition that will occur in photo-induced reactions. Broadly speaking, two types of non-adiabatic methods dominate the 
literature: Ehrenfest Dynamics (ED) methods and surface hopping (SH) methods. \cite{
Doltsinis-B2002, Barbatti2011, CHUCHORD2013} In the ED approach, the nuclei move classically on a single effective PES obtained by averaging over all the adiabatic states involved; however, in SH methods the classical degrees of freedom (nuclei coordinates) evolve on single adiabatic surfaces,\cite{Doltsinis-B2002} and make probabilistic hops from one PES to another. Surface-Hopping methods present key advantages over ED methods, because the system is always on a certain pure adiabatic state, not in an average non-physical state. Moreover, ED calculations sometimes lead to non-physical features, such as energetically inaccessible levels populated or violation of microscopic reversibility. \cite{Doltsinis2002-1}  

Photo-induced processes in large molecular systems are computationally intensive to simulate due to the complexities in modelling the vast search space associated with photo-activated reaction mechanisms. Also, the search for conical intersections is quite more complex than geometry optimization. \cite{Keal2007,Bearpark1994,Manaa1993,Yarkony2004}  Photo-induced polymerization reactions, such as the photo-induced [2+2] cycloaddition reactions of two C$_{60}$ molecules, would easily require large simulation systems containing hundreds of atoms. Biological photo-induced reactions are even more demanding in computational resources. \cite{Rossle2010} Additionally, photo-induced reaction mechanisms are probability events. Hence, to accurately simulate reaction mechanisms, hundreds, if not thousands, of trajectories 
are required to obtain the proper statistical information. Thus, efficient algorithms are needed, at least, at an exploratory level. There is a definitive need for simulation tools that can, at minimum, obtain sufficiently accurate results that will efficiently yield significant insight in photo-induced reaction mechanisms without overburdening computational resources. 

Tully's molecular dynamics with electronic transitions (MDET) algorithm\cite{Tully1990} employes a surface hopping strategy\cite{Tully1971} to switch between different adiabatic states. In Tully's MDET approach, the system evolves on one of the PES corresponding to a particular adiabatic state $E_A$ of the system.  Using the idea of ensemble of trajectories, the transitions between different PES are determined through a probabilistic 
approach; in particular, the transitions between adiabatic electronic states are determined using the  {\it fewest-switches algorithm} that considers the minimum number of non-adiabatic transitions that are compatible with the correct statistical distribution of state populations at all times.  
The MDET method is probably the most used and successful method to deal with problems in which coupling between electrons and ions is required.

The most efficient first-principles molecular dynamics (MD) approaches are based on density functional theory (DFT). In recent years the MDET approach has been implemented using DFT and applied to several problems. 
\cite{Doltsinis2002-1, Doltsinis2002-2,Craig2005,Duncan2008,Habenicht2008,Tapavicza2007,Tavernelli2009,Tapavicza2013,Neukirch2014,Mitric2009,Lan2011,Wohlgemuth2011,Tapavizca2011,Malis2012} 
Our goal is to target large-scale photo-polymerization trajectory ensembles that include several hundreds of atoms and explore several hundreds of trajectories. As such, we have focused primarily on implementing a MDET algorithm within {\sc Fireball}, an efficient local-orbital DFT MD technique.\cite{Lewis2011} Our approach to the nonadiabatic MD simulaton is extremely efficient because we pre-calculate and store the integrals needed during the MD simulation ({\it e.g.} the integrals for the calculation of the hamiltonian matrix elements).\cite{Sankey1989,Jelinek2005} During the simulations, interactions are loaded into memory from the pre-computed data files and no additional time is needed for {\it on-the-fly} integration. We have previously demonstrated a successful application of the {\sc Fireball} DFT-MDET in the photo-isomerization processes of stilbene and azobenzene.\cite{Neukirch2014} 

The remainder of this paper is organized as follows. In Section II, we present a detailed description of our implementation of DFT-MDET within the {\sc Fireball} code. In Section III, we present results for two different photo-induced cycloaddition reactions: the [2+2] cycloaddition of ethylene 
with maleic anhydride to form cyclobutane-1-2-dicarboxilic anhydride and the [2+2] photo-induced polymerization reaction of two C$_{60}$ molecules. Finally, in section IV, we summarize the main conclusions.

\section{Description of the Method}\label{sec:mdet}
In DFT-MDET simulations the atoms follow classical trajectories $\{\textbf{R}_{\alpha}(t)\}$ on one of the PESs corresponding to a particular adiabatic state $E_A$. The coupling between the nuclear motion and the electronic quantum state is taken into account by means of probabilistic hops between different adiabatic PESs ($E_A \rightarrow E_B$). The probability for these hops is determined from the time evolution of the electronic states and their non-adiabatic coupling vectors (see below). In the following discussion, we present details about our implementation of the MDET algorithm within the {\sc Fireball} local-orbital DFT code.

\begin{figure}[h!]
  \centering
    \includegraphics[width=0.5\textwidth]{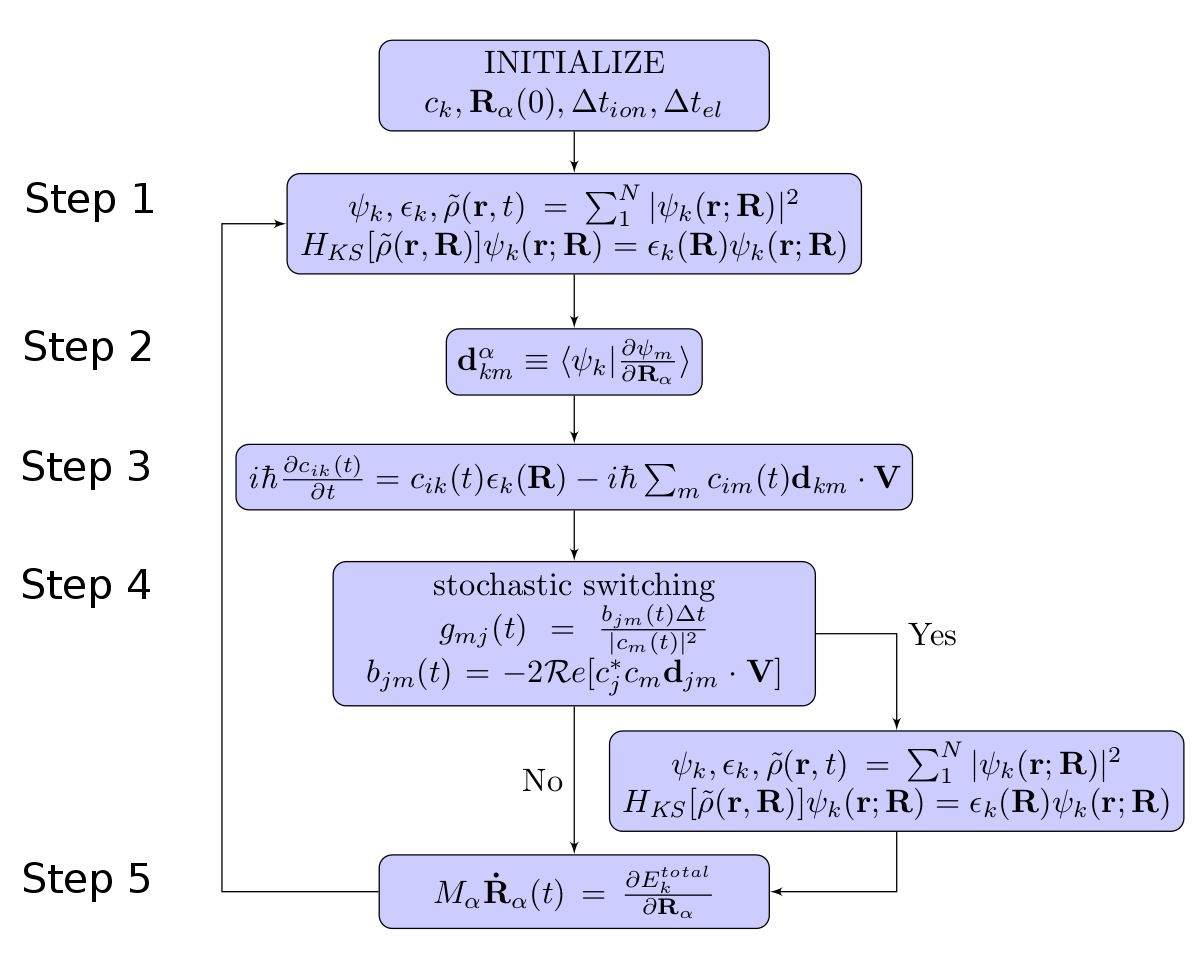}
  \caption{
  The MDET simulation consists of two nested time loops. Step 1 corresponds to the self-consistent solution of the stationary KS Hamiltonian for the current PES at time $t$; in step 2 the non-adiabatic coupling term between adiabatic KS states are calculated; step 3 is the inner loop for the time evolution of the electronic states; step 4 corresponds to the fewest switches algorithm: surface hops between different PES are determined using the probabilities $g_{mj}$; if there is a hop a new self-consistent solution is obtained for the KS Hamiltonian for the new electronic configuration and energy is conserved by means of velocity rescaling along the direction of the NACV; finally in step 5 the atomic positions are updated $t \rightarrow t + \Delta t$ (outer loop).
  }
  \label{fig:scheme}
\end{figure}

The MDET simulation consists of two nested time loops corresponding to the time evolution of the atomic positions and electronic states; the outer time loop corresponds to the atomic motion and the inner time loop to the propagation of the electronic states. The computational procedure consists of 5 algorithm steps (after an initialization step) as shown schematically in \reffig{scheme}.

{\bf Step 1: Solution of the stationary Kohn-Sham Hamiltonian.} 
In this first step we obtain the self-consistent (SCF) solution of the stationary Kohn-Sham (KS) Hamiltonian for the current atomic configuration and adiabatic state:

\begin{equation}\label{eq:statKS}
\hat{H}_{\mathit{KS}}[\mathbf{R}] \psi _{j}(\mathbf{r};\mathbf{R}) = \epsilon_{j}(\mathbf{R})\psi_{j}(\mathbf{r};\mathbf{R}),
\end{equation}
where $\mathbf{R}$ represents the atomic positions 
$\{\textbf{R}_{\alpha}(t)\}$ at time $t$ and $\psi _{j}(\mathbf{r};\mathbf{R})$ are the single-particle adiabatic KS orbitals (at time $t$). 
The different adiabatic states, $E_A$, are defined by means of the corresponding set of occupied KS adiabatic orbitals or, in other words, by the occupancies $f_k$ of the $\psi _{k}$ orbitals, with $f_k = 1$
(occupied) or $0$ (empty). 
Thus, the KS Hamiltonian in \refeq{statKS} is defined using the electronic configuration 
for the current PES ({\it e.g.} the electron density is
$\rho =\sum_k f_k \lvert \psi_k \lvert ^2$); for this purpose, we perform a {\it constrained} DFT calculation 
in which the level occupancy is imposed for a particular PES 
to maintain 
the electronic configuration during the SCF cycle.  In order to keep track of the identity of the adiabatic KS states at different time steps 
we calculate the overlaps $\langle \psi _{k} (t - \Delta t) | \psi _{m} (t) \rangle$.

{\bf Step 2: Calculation of the non-adiabatic couplings.}
The time evolution of the electronic state is calculated using time-dependent KS theory. \cite{Runge84,Marques04,Prezhdo2005} For this purpose, we expand the time-dependent KS orbitals, $\varphi_i$, in terms of the adiabatic KS states 
\begin{equation}\label{eq:dadtKS}
\varphi_i(\mathbf{r},t)=\sum _{k}c_{ik}(t)\psi_{k}(\mathbf{r};\mathbf{R}),
\end{equation}
and evolve them using the time-dependent, Schr\"odinger-like, KS equations:
\begin{equation}\label{eq:H_KS}
\hat{H}_{\mathit{KS}}[\mathbf{R}]\varphi_i(\mathbf{r},t) = i\hbar \frac{\partial \varphi_i (\mathbf{r},t)}{\partial t}.
\end{equation}
This yields the equations of motion for the coefficients $c_{ik}(t)$ in \refeq{dadtKS}:
\begin{equation}\label{eq:teKS}
\mathit{i\hbar}\frac{\partial c_{ik}(t)}{\partial t}=c_{ik}(t)\epsilon_{k}(\mathbf{R})-i\hbar \sum_{m}c_{im}(t)\mathbf{d}_{\mathit{km}}\cdot \mathbf{V}.
\end{equation}
In this equation the coupling between the classical motion of the atoms and the time evolution of the electronic quantum state is reflected in the non-adiabatic coupling (NAC) term:
\begin{equation}
\textbf{d}_{km} \cdot  \textbf{V} \equiv
\sum_\alpha \ { \textbf{d}_{km}^\alpha \textbf{V}_{\alpha} },
\end{equation}
where $\textbf{V}_{\alpha} = {\partial \textbf{R}_{\alpha}}/{\partial t}$ is the atomic velocity of atom $\alpha$
and  $\textbf{d}_{km}^{\alpha}$ are the non-adiabatic coupling vectors (NACV) between adiabatic KS states $k$ and $m$:
\begin{equation}\label{eq:nacKS}
\mathbf{d}_{\mathit{km}}^{\alpha }\equiv \langle \psi _{k}|\frac{\partial \psi _{m}}{\partial \mathbf{R_{\alpha}}}\rangle.
\end{equation}
The NACVs $\textbf{d}_{km}^{\alpha}$ are the central quantity for non-adiabatic molecular dynamics. The atomic motion induces changes in the adiabatic states populations $| c_{ik}(t) |^2$ through the coupling of the atomic velocities and the NACVs, see \refeq{teKS}.
Recently, we have derived an expression to calculate the NACVs $\textbf{d}_{km}^{\alpha}$ in a basis set of local-orbitals {\it on the fly} along the MD simulation in a practical and computationally efficient way.\cite{Abad2013} 
Alternatively, the NAC for the time evolution (\refeq{teKS}) of the coefficients $c_{ik}(t)$  can be directly calculated using the following numerical derivative \cite{Hammes-schiffer1994}:
\begin{equation}
 \textbf{d}_{km} \cdot  \textbf{V} \equiv 
 \langle \psi _{k}|\frac{\partial \psi _{m}}{\partial t} \rangle = 
 \frac{1}{2 \Delta t} 
 \left[
 \langle \psi _{k} (t - \Delta t) | \psi _{m} (t) \rangle -
 \langle \psi _{k} (t) | \psi _{m} (t - \Delta t) \rangle
 \right],
\end{equation}
which gives the value of the NAC at time $(t - \Delta t/ 2)$. 

{\bf Step 3: Time evolution of the electronic states.} 
Once the NACs have been calculated, we propagate the coefficients $c_{ik}$ from time $(t - \Delta t)$ to time $t$ using a small time step $\Delta t_{el}$ and a 4-th order Runge-Kutta numerical scheme to integrate \refeq{teKS}. Typically 
$\Delta t_{el} \sim \Delta t / 500$. 
This is the inner loop for the propagation of the electronic states for each step, $ \Delta t $,  in the atomic motion.

{\bf Step 4: Surface hopping.}
At each time step $\Delta t$, a transition between adiabatic PESs $E_A \rightarrow E_B$ 
may occur through a probabilistic ``hop" using the  {\it fewest-switches algorithm}.\cite{Tully1990} The MDET algorithm incorporates the following  switching probability $g_{jk}(t)$ to define the transition between electronic states $ \psi_j \rightarrow \psi_k$ in the time interval $ [t - \Delta t, t]$:
\begin{equation}\label{eq:fews}
g_{jk} (t) = \frac{ b_{kj} (t) \Delta t}{|c_{jj} (t)|^2},
\end{equation}
where the coefficient $b_{kj}$ is defined as:
\begin{equation}
b_{kj} (t) = -2 {\cal R}e [ c_{jk}^{*}(t) c_{jj}(t) \textbf{d}_{kj} \cdot \textbf{V} ].
\end{equation}
The transitions are determined using a Monte-Carlo approach following the probabilities $g_{jk}(t)$. When a transition takes place, the occupancies of the adiabatic KS orbitals $ \psi_j$ and $\psi_k$ are modified:
$f_j: 1 \rightarrow 0; f_k: 0 \rightarrow 1$, and the stationary KS Hamiltonian, \refeq{statKS}, is re-calculated for the new electronic configuration (see step 1). In this case, energy conservation is imposed by re-scaling velocities along the direction of the NACV, \refeq{nacKS}).\cite{Coker1995,Herman1985} Hops to higher energy PESs are rejected if insufficient kinetic energy is present. Velocity rescaling and hop rejection ensures detailed balance between transitions up and down in energy.
If the switch does not occur, the simulation continues on the current PES. 
 
{\bf Step 5: Motion of the atoms.} 
Finally, the atomic positions are updated
$\{\textbf{R}_{\alpha}(t)\} \rightarrow \{\textbf{R}_{\alpha}(t + \Delta t)\}$ using the forces corresponding to the current PES, $E_A$:
\begin{equation}\label{eq:forces}
\textbf{F}_{\alpha} = - \frac{\partial E_{A}}{\partial \textbf{R}_{\alpha}}; 
\end{equation}
where $\textbf{F}_{\alpha}$ is the force acting on atom $\alpha$. The calculation goes back to step 1, using the new atomic positions.  This scheme is repeated until the simulation is finished.

We now describe the parameters used during simulations of photo-cycloaddition (PCA) reactions discussed in the next section. The optimized atomic structure in the ground state is used as the initial molecular configuration for the reactant species. Atomic positions of the reactant species were relaxed with convergence criteria for the total energy and forces of $10^{-6}$ eV and $0.05$ eV/\AA\,  respectively to obtain the optimized structures. All calculations were performed using local density approximation (LDA). The optimized numerical 
atomic-like orbitals (NAOs) were confined to regions limited by the corresponding cutoff radii $r_c$.\cite{Sankey1989, Basanta2007} The following $r_c$ for the different NAOs were chosen as follows: hydrogen, $r^H_c(s)$= 3.8 a.u., with an additional 
$s^*$-state, $r^H_c(s^*)$= 3.8 a.u.;  carbon, $r^C_c(s)$= 4.5 a.u, $r^C_c(p)$= 4.5 a.u., with an additional 
$d$-state, $r^C_c(d)$=5.4 a.u.; and oxygen, 
$R^O_c(s)$= 3.8 a.u. and $R^O_c(p)$= 3.8 a.u. All molecular dynamics simulations were carried out with the characteristic time steps $t_{ion}$ = 0.25 fs 
and $t_{el}$ =0.0025 fs at temperature 500K and 100K for 
maleic anhydride and C$_{60}$, respectively. Constant temperature molecular dynamics was achieved using an isokinetic NVT thermostat.\cite{Evans1983}  

\section{Results: Cycloaddition}
\subsection{Cycloaddition reaction of two small organic molecules: 
reaction of maleic anhydride with ethylene.}
\begin{figure}[h!]
  \centering
    \includegraphics[width=0.8\textwidth]{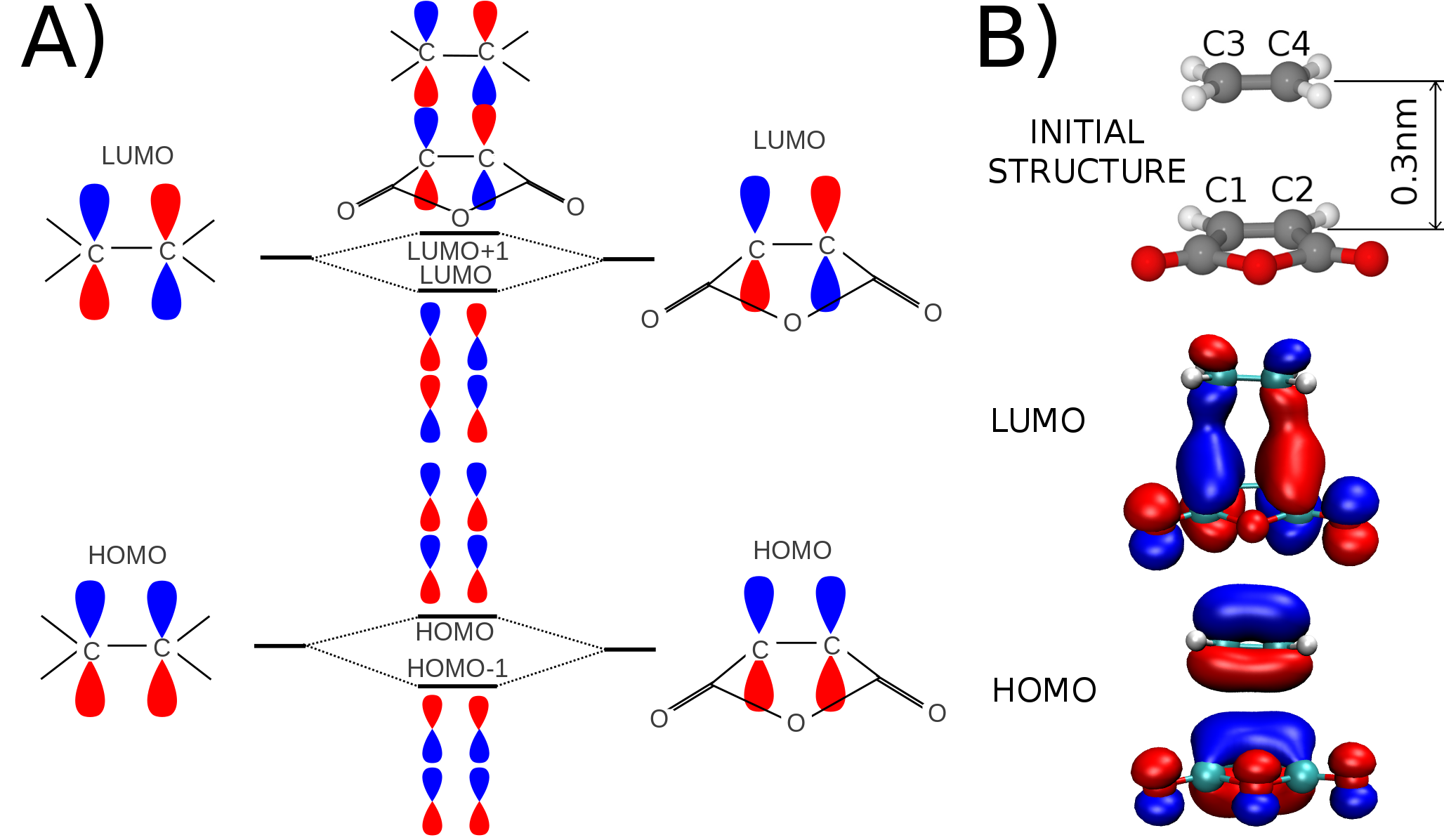}
      \caption{ A) Schematic representation of the frontier molecular orbitals involved in the PCA reaction. The figure shows the $p_z$ orbitals associated with the carbon-carbon double bonds in ethylene and maleic anydride. B) (Top) The initial configuration of the reactants. Molecular planes are parallel and displaced 0.3 nm. B) (Bottom) The HOMO and LUMO Kohn -Sham orbitals for the molecular complex in the initial configuration depicted in real space, showing the bonding and anti-bonding character of the LUMO and HOMO, respectively. Red and blue color express positive or negative sign of the wavefunction. }
  \label{fig:anh-0}
\end{figure}

To demonstrate the efficiency of our method, we first report DFT-MDET simulations of the photo-cycloaddition (PCA) reaction of maleic anhydride ($C_2O_3H_2$) with ethylene ($C_2H_4$), forming cyclobutane-1-2-dicarboxilic anhydride ($C_4O_3H_6$), see \reffig{anh-0}B.  This PCA reaction is a~prototypical reaction frequently encountered in organic chemistry. The reaction is generated by a single electron excitation induced by the interaction with light.  During the PCA reaction the two $\pi$ bonds perish and two $\sigma$ bonds arise, thus 4 bonding electrons are involved in the reaction process. The frontier molecular orbital description of PCA is schematically presented in \reffig{anh-0}A, where the $p_z$ character of the frontier orbitals, the HOMO and LUMO of the isolated reactants, are displayed on the right and left sides of the figure. The formation of a~molecular complex by the two reactants leads to four linear combinations of the frontiers orbitals (shown in center of \reffig{anh-0}A).  Kohn-Sham HOMO 
and LUMO orbitals in real space for the 
molecular complex are shown in the bottom of 
\reffig{anh-0}B. The HOMO orbital is formed by the overlap of $\pi$ bonds of opposing signs, which gives rise to the anti-bonding character of the HOMO orbital, see \reffig{anh-0}A. Consequently, the anti-bonding character of the HOMO orbital inhibits a~thermally activated reaction.  On the contrary, the LUMO orbital shows bonding characteristics. 
Thus, a photo-induced electron excitation from the HOMO to the LUMO induces an additional attractive force which facilitates the PCA reaction, as discussed below.
 
To gain more insight into the reaction mechanisms, we performed, in total, 50 independent DFT-MDET simulations, each with total time of 2 ps; in each simulation, a single electron was initially promoted 
from the HOMO to the LUMO state of the molecular complex ($C_2H_4 + C_2O_3H_2$). Each simulation took approximately 16 hours on single CPU  Intel Xeon E5420 @ 2.50GHz which demonstrates the computational efficiency of our FIREBALL-MDET code.  All simulations were initiated from an identical configuration - both planar molecules were oriented parallel and the carbon-carbon double bond ($C_3 = C_4$)  of ethylene was positioned above the carbon-carbon double bond  ($C_1 = C_2$)  of maleic anhydride (see \reffig{anh-0}B top). The initial distance between the molecular planes of the anhydride and ethylene was 3 \AA. The velocities were initialized with a normal distribution corresponding to an initial temperature of 500 K. 
In order to show clearly the stochastic nature of the algorithm, the same initial positions and velocities were applied for each trajectory.  Four frontier 
orbitals (HOMO-1, HOMO, LUMO, LUMO+1) were propagated in time and thus the electron transitions were allowed only amongst these states. 

We were able to identify three different deactivation channels of the initial electron excitation according to our simulations: (i)  a~direct addition reaction of ethylene and the anhydride;  (ii) a~direct non-reactive de-excitation; and (iii) an~intermediate transient state with subsequent de-excitation leading to an addition/dissociation process. According to our simulations, 15, of a total 50, simulations produced the addition reaction forming cyclobutane diacid anhydride; thereby our simulations give an estimated yield of the addition reaction as $\approx$ 30 \%. 

\begin{figure}[h!]
 \centering
   \includegraphics[width=0.5\textwidth]{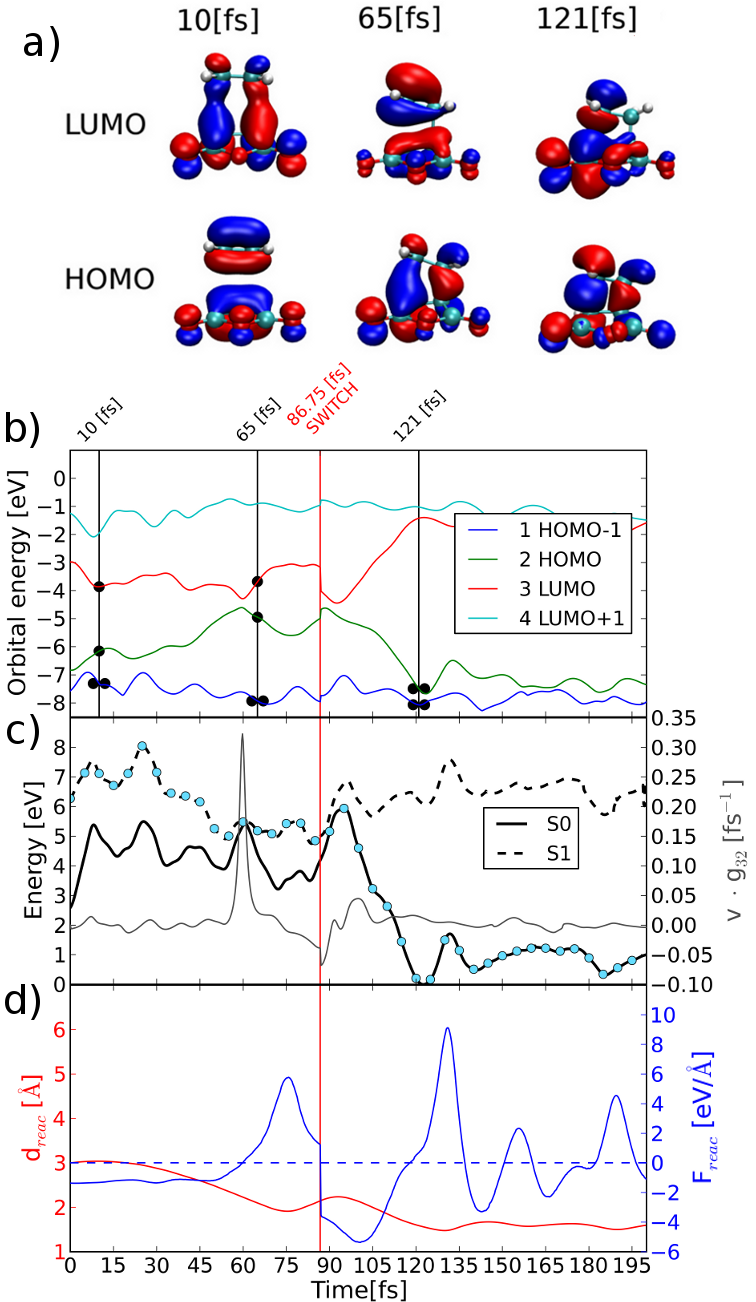}
 \caption{Analysis of a~direct addition reaction of ethylene and 
 maleic anhydride.  a) Character in real space of HOMO and LUMO electronic wave functions corresponding to the time points indicated by black vertical lines on graph b.
 b) Energy spectra for the four frontier molecular orbitals; the black dots indicate the electron occupancies of the different states. c) Potential energy surfaces corresponding to the ground S$_0$ and first excited S$_1$ state together with the nonadiabatic coupling term $\mathbf{d}_{ij} \cdot \mathbf{V}$ between HOMO and LUMO (solid grey line); the cyan dots indicate the actual PES for the simulation. d) Distance between the center of ethylene carbon-carbon bond ($C_1 - C_2$) to the center of maleic anhydride carbon-carbon bond ($C_3 - C_{4}$) (red) and forces acting between reactants (blue).}
 \label{fig:anh-1a}
\end{figure}

A direct addition reaction of reactants was observed in 12 cases. The detailed analysis of the reaction pathway leading to direct formation of cyclobutane diacid anhydride is presented in \reffig{anh-1a}. \reffig{anh-1a}a) represents the character in real space of the HOMO and LUMO electronic wave functions at different times along the simulations; in particular the HOMO and LUMO wave functions shown in \reffig{anh-1a}a) correspond to the time positions denoted by black vertical lines on \reffig{anh-1a} b).  The time evolution of the energy spectrum for the four frontiers molecular orbitals with their temporal occupancies is shown in \reffig{anh-1a} b). Complementary,  \reffig{anh-1a} c) shows the PES corresponding to the ground S$_0$ and first excited S$_1$ state along the simulation; the cyan dots indicate the actual PES for the simulation.  The grey line represents the evolution of the NACVs between the HOMO and the LUMO orbitals with its maximum occurring near the conical intersection on \
reffig{anh-1a} c).

To characterize the pathway of the 
reaction process, we traced a~time evolution of a~characteristic distance $ d_{reac}$ and force $F_{reac}$ acting between reactants. The characteristic distance $ d_{reac}$ measures the distance from the center of ethylene carbon-carbon bond ($C_1=C_2$, see \reffig{anh-0}) to the center of anhydride carbon-carbon bond ($C_3=C_4$). We also evaluated the forces acting between the two molecules  as $ F_{reac}=\sum {F_{ethylene}} - \sum {F_{anhydride}} $ projected along the vector
from the center of ethylene carbon-carbon bond to the center of anhydride carbon-carbon bond
($\vec{d}_{reac}$). The distance $d_{reac }$ and force $F_{reac}$ between reactants are plotted in \reffig{anh-1a} d) by the red and blue curves, respectively.

Initially, the HOMO and LUMO have anti-bonding and bonding characters, respectively. The partial occupancy  of the bonding state (LUMO) induces an attractive force acting between reactants. Consequently their distance $d_{reac}$ reduces significantly from 3.0 \AA\ to 2 \AA. Within $\approx$ 60 fs  the system approaches towards the avoided crossing region and the HOMO and LUMO  exchange their characters - the HOMO is now a bonding orbital and vice versa. 
In this particular case, after the ''crosing`` of the HOMO and LUMO, a~de-excitation of the electron from LUMO (anti-bonding) to HOMO (bonding) happens at approximately 86 fs, which stabilizes the final molecular complex and produces the addition reaction.  The force $F_{reac}$ becomes strongly attractive, driving the 
system to the final bound state 
with the distance $d_{reac}$ oscillating around 1.6 \AA. 

\begin{figure}[ht]
\centering
\includegraphics[width=0.5\textwidth]{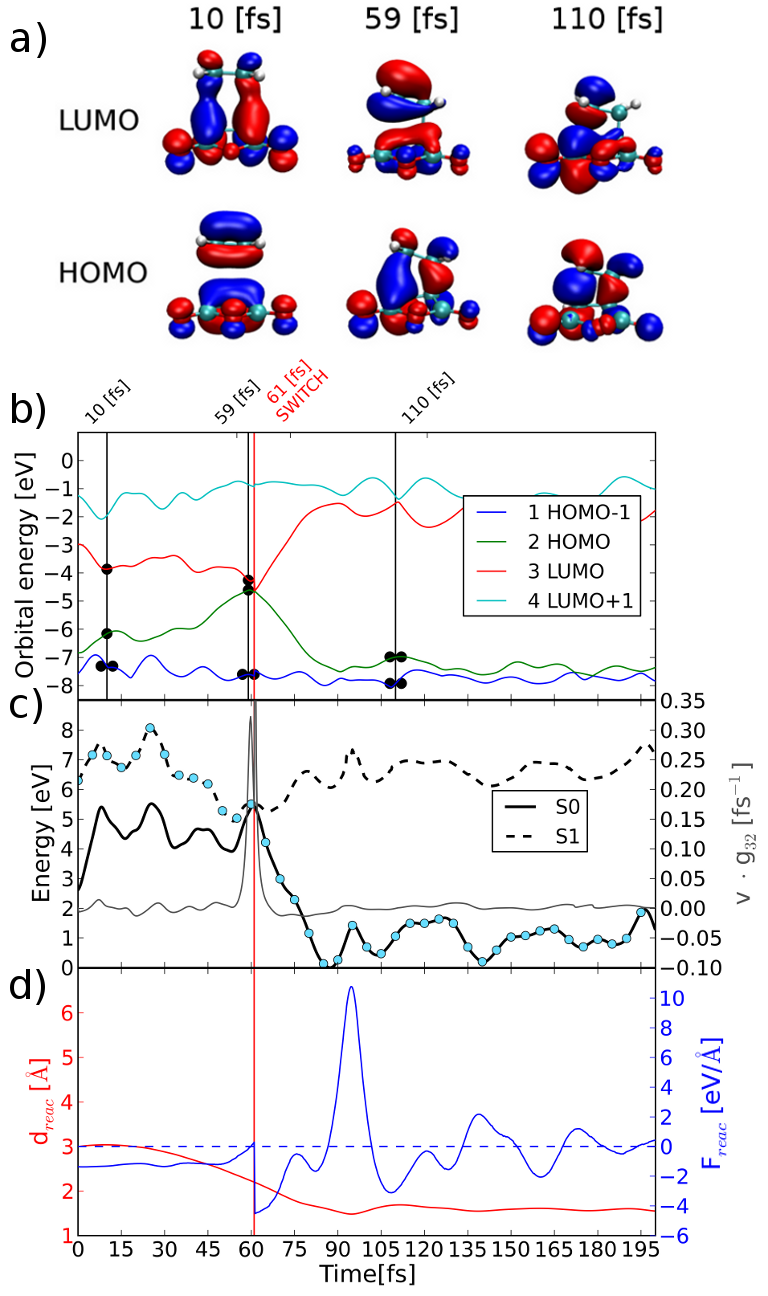}
\caption{ Analysis of a~direct addition reaction of ethylene and 
maleic anhydride with the electronic transition near the conical intersection.  
a) 
Character in real space of HOMO and LUMO electronic wave functions corresponding to the time points indicated by black vertical lines on graph b.
 b) Energy spectra of four frontier molecular orbitals; the black dots indicate the electron occupancies of the different states. 
 c) Potential energy surfaces corresponding to the ground S$_0$ and first excited S$_1$ state together with nonadiabatic coupling term $\mathbf{d}_{ij} \cdot \mathbf{V}$ between HOMO and LUMO (solid grey line); the cyan dots indicate the actual PES for the simulation. d) Distance between the center of ethylene carbon-carbon bond ($C_1 - C_2$) to the center of maleic anhydride carbon-carbon bond ($C_3 - C_{4}$) (red) and forces acting between reactants (blue).}
\label{fig:anh-1b}
\end{figure}

We should note that, in this particular case, the switching mechanism does not occur at the distance with the NACV maximum, but later due to the stochastic nature of the fewest switches MDET procedure.  Nevertheless, in other trajectories we also observe a LUMO to HOMO electron transition near the crossing region, as shown 
in \reffig{anh-1b}. Here the electronic transition from LUMO to HOMO happens at approximately 61 fs, very near to the conical intersection, with the HOMO 
exhibiting a bonding character. 
After this transition, the PCA reaction is completed.

\begin{figure}[ht]
\centering
\includegraphics[width=0.5\textwidth]{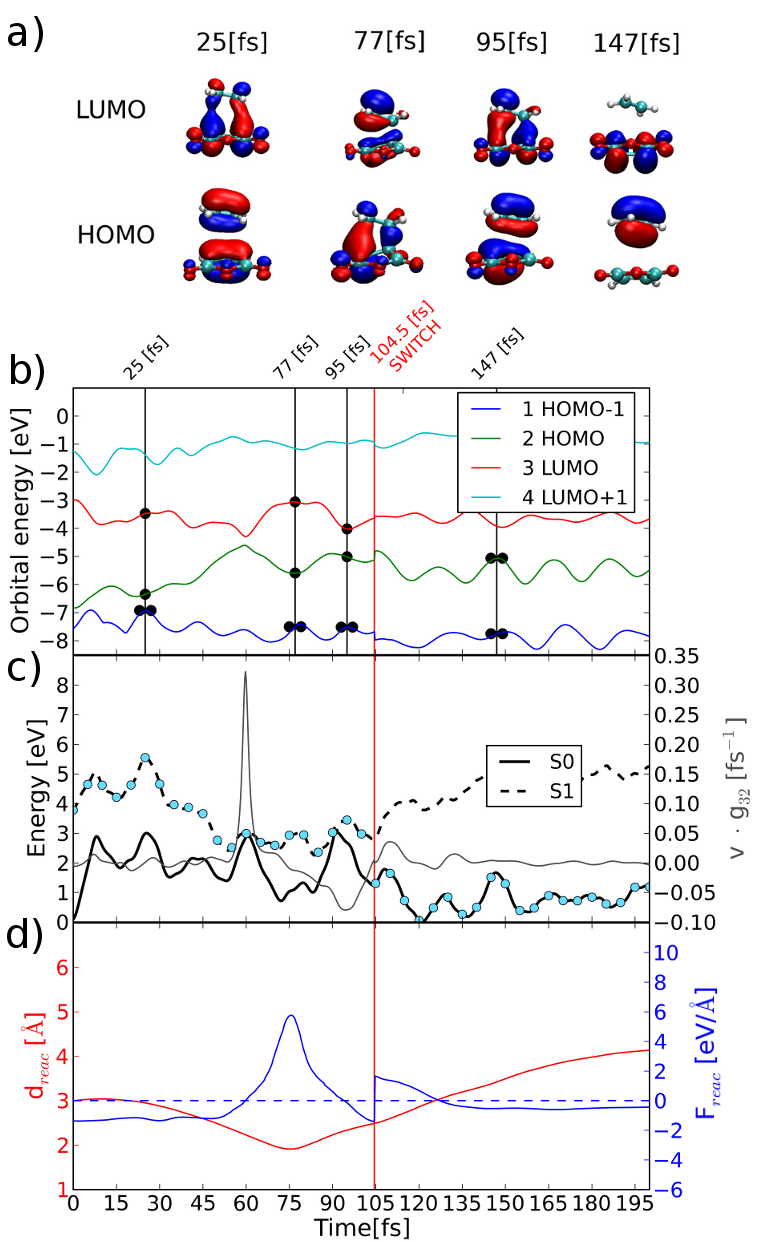}
\caption{ Analysis of a~direct non-reactive de-excitation of ethylene and 
maleic anydride.  
a) 
Character in real space of HOMO and LUMO electronic wave functions corresponding to the time points indicated by black vertical lines on graph b.
 b) Energy spectra of four frontier molecular orbitals; the black dots indicate the electron occupancies of the different states. 
 c) Potential energy surfaces corresponding to the ground S$_0$ and first excited S$_1$ state  together with nonadiabatic coupling term $\mathbf{d}_{ij} \cdot \mathbf{V}$ between HOMO and LUMO (solid grey line); the cyan dots indicate the actual PES for the simulation. d) Distance between the center of ethylene carbon-carbon bond ($C_1 - C_2$) to the center of maleic anhydride carbon-carbon bond ($C_3 - C_{4}$) (red) and forces acting between reactants (blue).}
\label{fig:anh-3}
\end{figure}

The second deactivation channel consists of the direct nonreactive de-excitation. In our simulations, we have detected 13 cases of the direct dissociation from the total 50. One characteristic simulation leading to the dissociation is shown in \reffig{anh-3}.  Initially the system evolves along the same trajectory as in the previous cases. It goes through the avoided crossing region, where the HOMO and LUMO orbitals exchange their position keeping each one electron as shown in \reffig{anh-3}a). No electron hopping occurs this time and the systems evolves towards a second avoided crossing region, which reaches at time $\approx$ 95 fs. Here the HOMO and LUMO orbitals exchange their position again, now the HOMO and LUMO having anti-bonding and bonding characteristics, respectively, similar to the initial configuration.  The electron hopping occurs between the LUMO and the HOMO at 104 fs, which leads to a double occupancy of the anti-bonding HOMO state;  
this induces an important change in the force between both molecules, that after the transition presents a large repulsive value (~2 eV/\AA ), driving the molecules far from each other (see \reffig{anh-3}d). Thus, the addition reaction is not produced.

\begin{figure}[h!]
 \centering
   \includegraphics[width=0.5\textwidth]{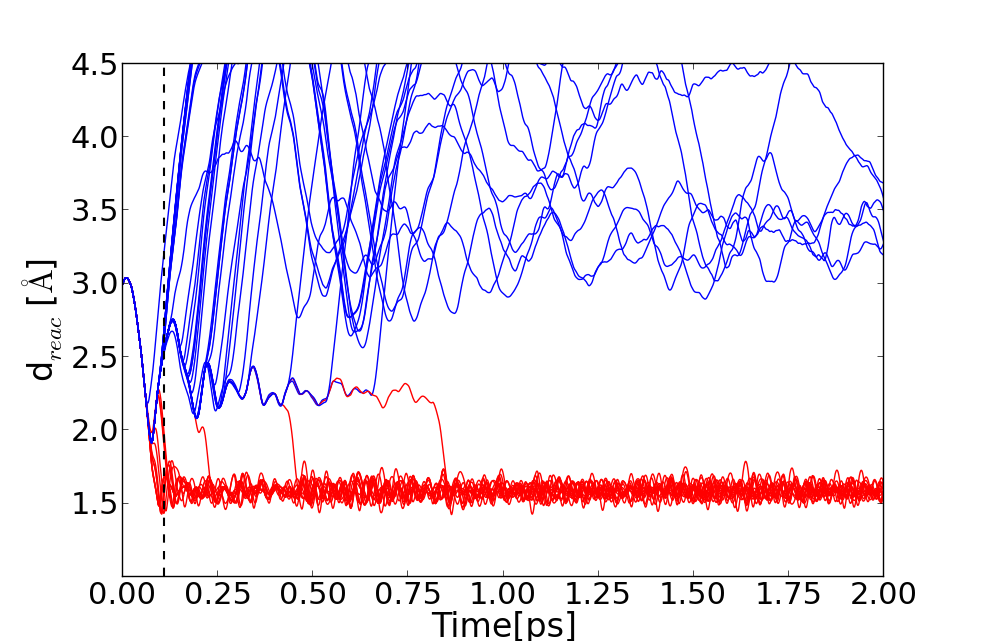}
 \caption{Time evolution of the characteristic distance $d_{reac}$ measured between the center of ethylene carbon carbon bond ($C_1 - C_2$) to the center of anhydride carbon carbon bond ($C_3 - C_4$). Red curves represent case when the addition reaction occurred and the blue curves represent the dissociation process.}
 \label{fig:anh-distance}
\end{figure}

\begin{figure}[ht]
\centering
\includegraphics[width=0.5\textwidth]{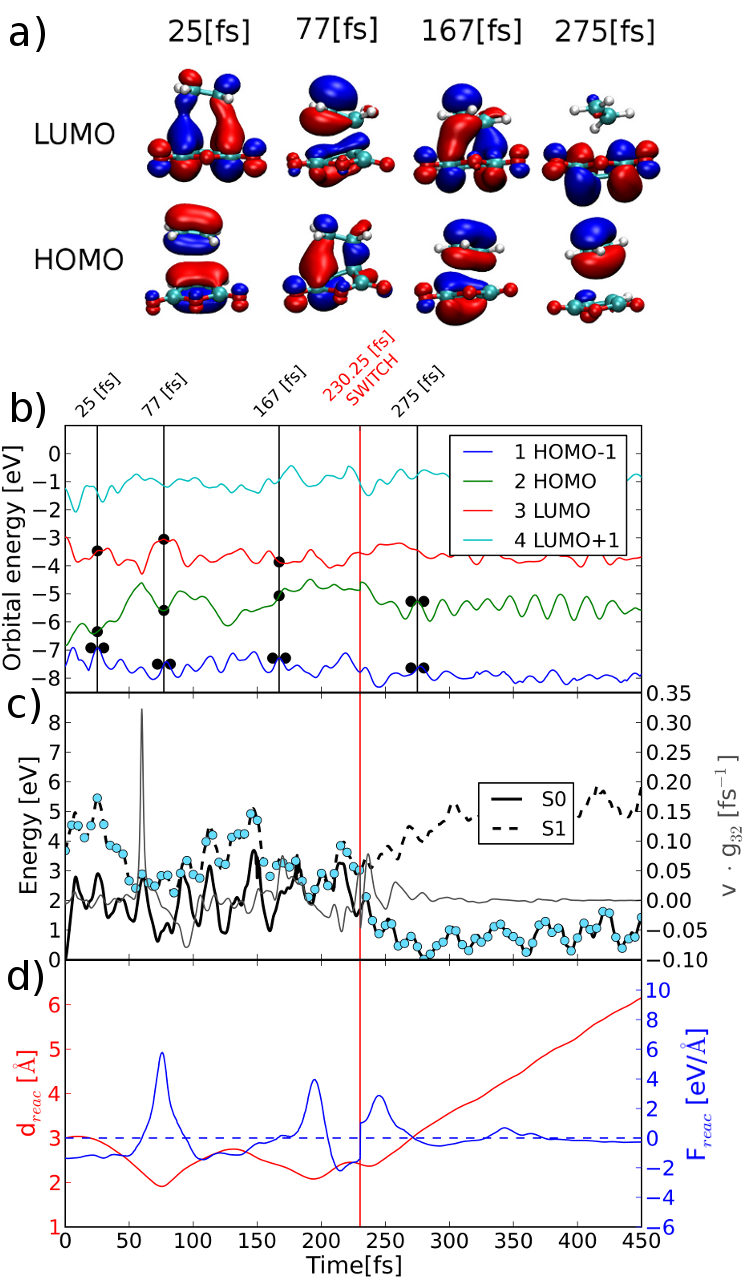}
\caption{ Analysis of the an~intermediate transient state with subsequent de-excitation leading to dissociation process of ethylene and maleic anydride.  
a) Character in real space of HOMO and LUMO electronic wave functions corresponding to the time points indicated by black vertical lines on graph b.
 b) Energy spectra of four frontier molecular orbitals; the black dots indicate the electron occupancies of the different states. 
 c) Potential energy surfaces corresponding to the ground S$_0$ and first excited S$_1$ state  together with nonadiabatic coupling term $\mathbf{d}_{ij} \cdot \mathbf{V}$ between HOMO and LUMO (solid grey line); the cyan dots indicate the actual PES for the simulation. d) Distance between the center of ethylene carbon-carbon bond ($C_1 - C_2$) to the center of maleic anhydride carbon-carbon bond ($C_3 - C_{4}$) (red) and forces acting between reactants (blue).}
\label{fig:anh-2}
\end{figure}

The third deactivation channel is related with the formation of an intermediate state which is characterized by the formation of just one carbon-carbon bond between maleic anhydride and ethylene. In this intermediate state the distance $d_{reac}$ oscillates around 2.3 \AA\ as shown on \reffig{anh-distance} and \reffig{anh-2}.  
In this case the system typically finalizes (22 cases from 50 total) in the decoupling of the molecular reactants but we detected 3 cases where the intermediate case produces the complete addition reaction (see \reffig{anh-distance}).  One typical example of the transition state, which evolves into the full dissociation is presented in \reffig{anh-2}. At the beginning of this transition state reaction profile, the time evolution of the system is the same as in the previous dissociative cases 
(e.g. see \reffig{anh-3}). However, the system undergoes two avoided crossing regions without any electron hopping. Therefore, at times larger than 150 fs the HOMO is occupied by one electron and it possesses an anti-bonding character. 
The system spends a relatively long time in the partial occupancy of the bonding LUMO orbital leading to the formation of a partial bonding state between two carbon atoms of ethylene and the anhydride, which introduces an asymmetric position of the ethylene molecule above the anhydride molecule (see \reffig{anh-2} a). Both the HOMO and LUMO become more coupled but still retaining their original bonding (LUMO) and anti-bonding (HOMO) characters as seen from \reffig{anh-2}a) at the time of 167 fs. The system remains in the intermediate state for about 100 fs with a relatively large (and oscillating) value of the NACV between HOMO and LUMO. The de-excitation occurs at time of 230 fs leading to a full occupancy of an anti-bonding HOMO. As a result, the ethylene molecule dissociates from the anhydride as seen from the increase of the distance $ d_{reac}$ plotted on \reffig{anh-2}d). For the rest of the simulation, the system is in the ground state and the HOMO orbital is localized 
on the ethylene molecule.

\subsection{Polymerization of two $C_{60}$ molecules via cycloaddition}
\begin{figure}[h!]
  \centering
    \includegraphics[width=0.8\textwidth]{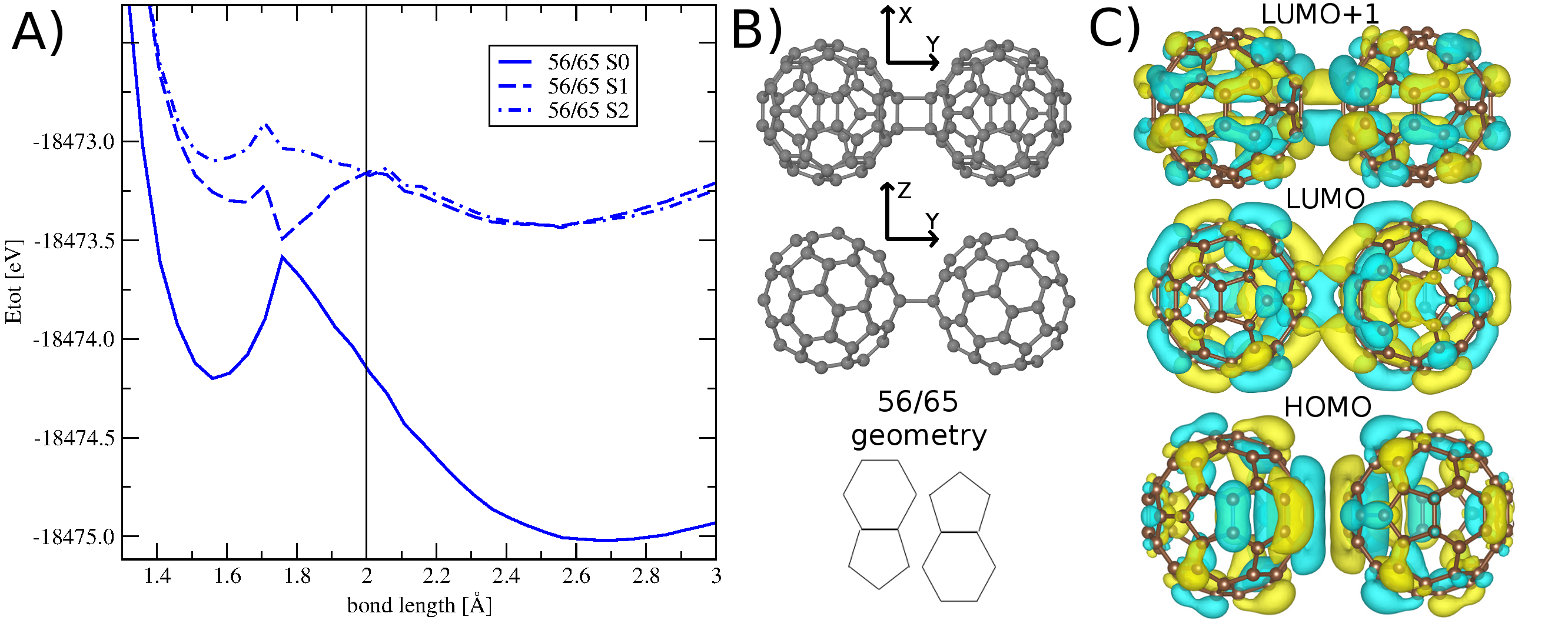}
  \caption{A) Potential energy surfaces of the PCA reaction between 2 fullerenes: 
  ground state (solid line), first-excited state (dashed line) and second excited state (dot-dashed line). 
  B) Atomic structure of the fullerene dimer of 56/65 geometry. 
  C) Real space representation of HOMO, LUMO and LUMO+1 KS electronic wave functions. While HOMO orbital has anti-bonding character, LUMO+1 is bonding.}
  \label{fig:C60-PES}
\end{figure}

Here, as a second example of the application of our FIREBALL-MDET code, we present simulations of the polymerization process of  two free-standing $C_{60}$ molecules via a 2+2 PCA reaction. This reaction was first reported by Rao {\it et al.} in 1993.\cite{Rao1993} The basic principles for this case are similar to our discussion in the previous section, see \reffig{anh-0} Accordingly, the HOMO of the molecular complex $C_{60}$-$C_{60}$ has an anti-bonding character and the bonding orbital is associated to the LUMO+1 (see \reffig{C60-PES}C). We should note that the LUMO and LUMO+1 orbitals are nearly degenerate with an energy difference of 0.08 eV in the initial configuration. Nevertheless the LUMO+1 shows a strong bonding character with pronounced electron densities localized between the frontiers atoms of the two $C_{60}$ molecules.  

Here we consider the case where two $C_{60}$ molecules are oriented by 56/65 edges ( see \reffig{C60-PES} B) against each other. \cite{Strout1993}  In the first step, we examine the PESs as function of distance between the two $C_{60}$   molecules in the ground $S_0$, first excited $S_1$, and second excited $S_2$ states for distances in the range from 1.3 to 3.0 {\AA}, with incremental changes in the distance of 0.05 \AA. 
In this calculations we fixed $z$ and $y$ coordinates of the four interacting carbon atoms forming the 56/65 edge see \reffig{C60-PES} B), {\it i.e.} those involved in the cycloaddition chemical bond, and the rest of the atomic positions were allowed to relax. The calculated PES S$_0$, S$_1$ and S$_2$ are plotted in \reffig{C60-PES} A). We can identify two minima on the S$_0$ PES: (i) a global one located at the distance of $\approx$ 2.69 \AA,  which corresponds to two 
weakly interacting $C_{60}$ molecules; (ii) and the second, a local minimum found at the distance of $\approx$ 1.56 \AA. This second minimum corresponds  to two chemically bonded $C_{60}$  molecules. In between
these two energy minima, there is an~energy barrier of $\approx$ 0.6 eV located at a distance of 1.75 \AA, which avoids a~spontaneous transition from the energetically higher chemically bound state to the weakly bound state. At this distance there is a conical intersection between S$_0$ and S$_1$.

In our DFT-MDET simulations, we initially placed two $C_{60}$ molecules at a distance of 2.0 \AA\  (vertical line in \reffig{C60-PES} A) {\it i.e.} 0.25 {\AA} away from the conical intersection, and one electron was promoted from the HOMO to the LUMO+1 state. During the simulation, we let the system evolve in time for 0.5 ps at a constant temperature of 100K, with initial velocities chosen as discussed in the previous section. Electron transitions are allowed between three orbitals - HOMO, LUMO and LUMO+1. Our simulations identify two scenarios: (i) a PCA reaction forming a $C_{60}$ 
bonded dimer; or (ii) a complete decoupling of the two $C_{60}$ molecules towards the weakly bound minimum. In total, we carried 20 trajectories and 3 of the simulations yielded the polymerization process; 17 simulations lead to the separation of the $C_{60}$ molecules. 
 
\begin{figure}[h!]
  \centering
    \includegraphics[width=0.5\textwidth]{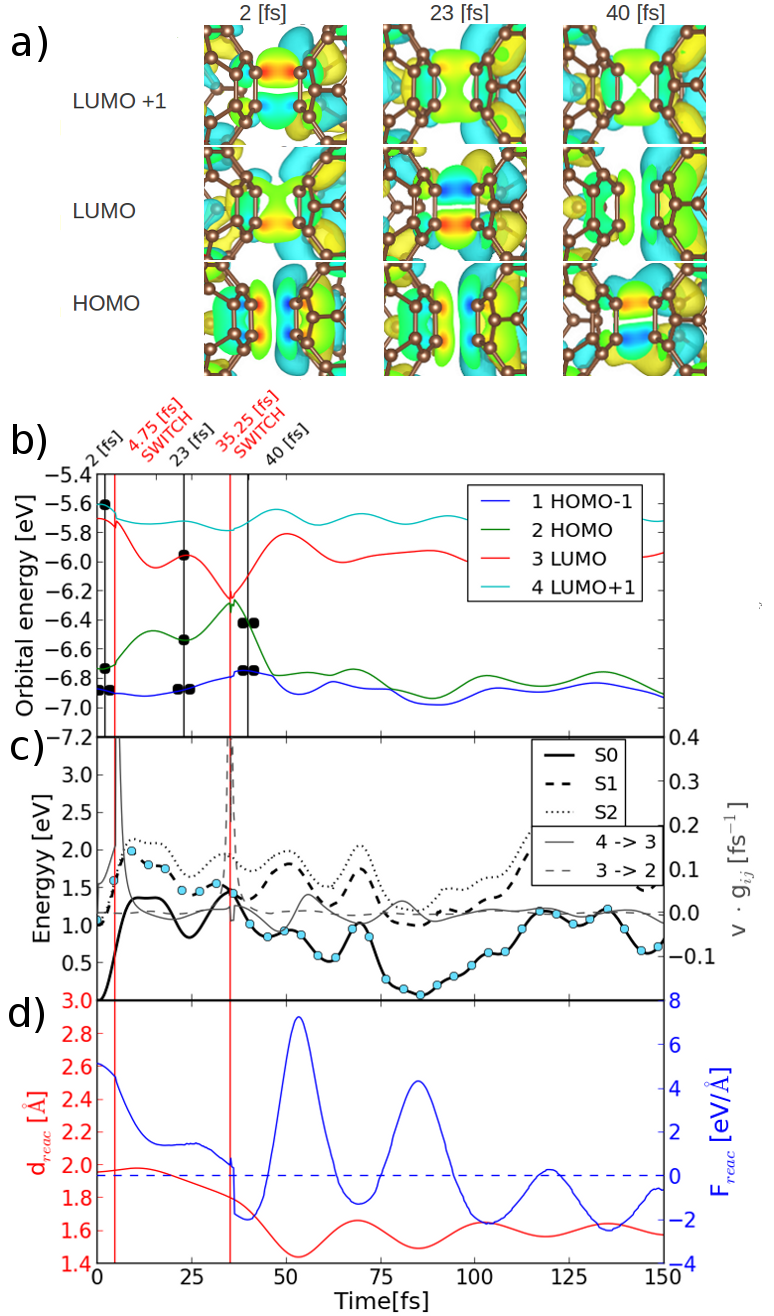}
  \caption{ a) Real space representation of HOMO, LUMO and LUMO+1 KS orbitals corresponding to the time points indicated by the black vertical lines on graph b. b) Energy spectra of four frontier molecular orbitals. c) Potential energy surfaces corresponding to the S$_0$, S$_1$ and S$_2$ states  together with the nonadiabatic coupling terms $\mathbf{d}_{ij} \cdot \mathbf{V}$ (solid grey line: between LUMO+1 and LUMO, dashed grey line: between LUMO and HOMO); the cyan dots indicate the actual PES for the simulation. d) Distance between the center of two reacting carbon atoms of the first fullerene to the center of the two reacting carbon atoms of the second fullerene (red) and forces acting between both molecules  (blue).}
  \label{fig:C60-bond}
\end{figure}


\begin{figure}[h!]
  \centering
    \includegraphics[width=0.5\textwidth]{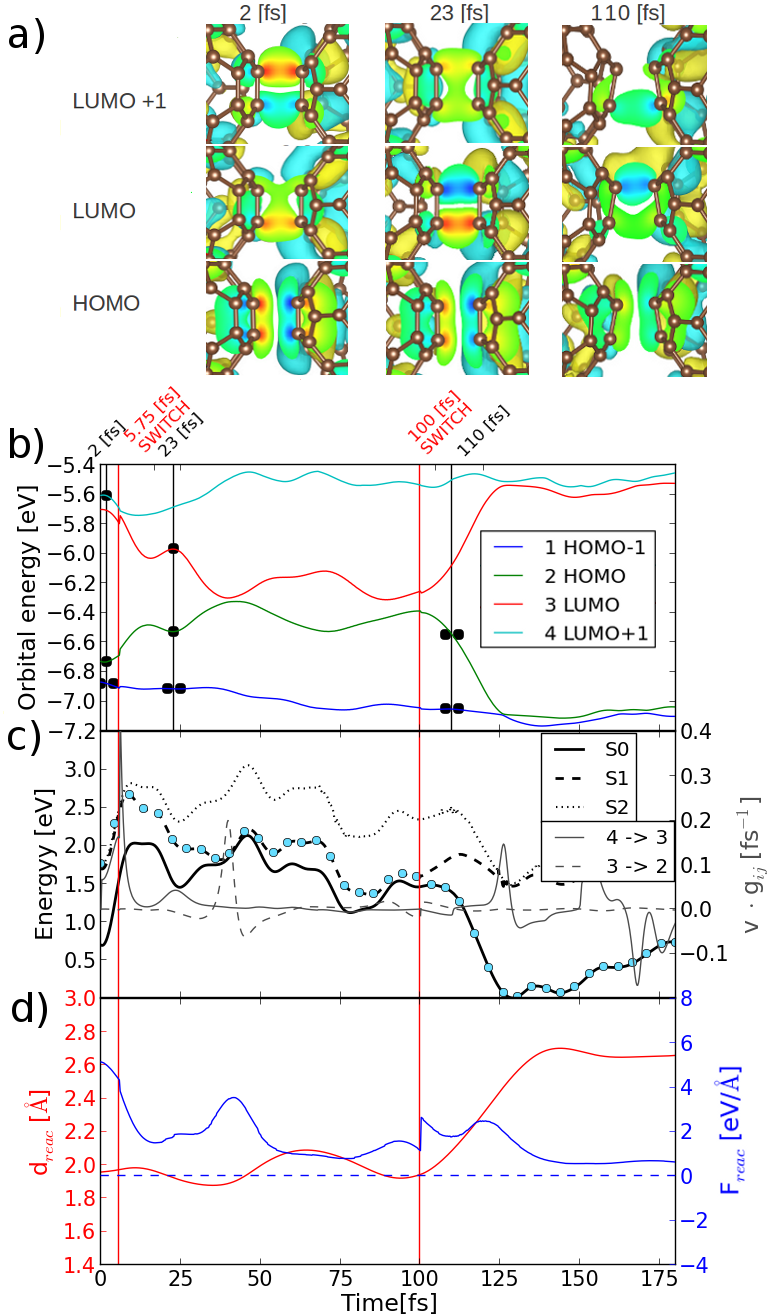}
  \caption{a) Real space representation of HOMO, LUMO and LUMO+1 KS orbitals corresponding to the time points indicated by the black vertical lines on graph b. b) Energy spectra of four frontier molecular orbitals. c) Potential energy surfaces corresponding to the S$_0$, S$_1$ and S$_2$ states  together with the nonadiabatic coupling terms $\mathbf{d}_{ij} \cdot \mathbf{V}$ (solid grey line: between LUMO+1 and LUMO, dashed grey line: between LUMO and HOMO); the cyan dots indicate the actual PES for the simulation. d) Distance between the center of two reacting carbon atoms of the first fullerene to the center of the two reacting carbon atoms of the second fullerene (red) and forces acting between both molecules  (blue).}
  \label{fig:C60-dis}
\end{figure}

\reffig{C60-bond} shows the time evolution for a~case where electron de-excitation happens via the 2+2 cycloaddition reaction.  \reffig{C60-bond} a) represents the HOMO, LUMO and LUMO+1 KS orbitals at three different snapshots corresponding to the vertical black lines in \reffig{C60-bond} b). \reffig{C60-bond} b) shows the time evolution of the HOMO, LUMO and LUMO+1 KS eigenvalues. \reffig{C60-bond} c) shows the PESs corresponding to the ground S$_0$, first S$_1$ and second S$_2$ excited states throughout the simulation. The cyan circles identify on which PES the system is located at a given time during the simulation.  The solid/dashed grey line represents the evolution of the NACVs between the LUMO and HOMO and between the LUMO+1 and LUMO, respectively.  Finally, \reffig{C60-bond} d) displays the evolution of the force $F_{reac}$ acting between the two molecules and their distance $d_{reac}$, which is defined as an~average distance between the 
frontiers atoms 
on the 56/65 edges of each molecule. Similarly, the force $ F_{reac}$ is the difference between the sum of forces acting on 
each $C_{60}$ molecule projected along the vector $d_{reac}$.

Initially, the S$_1$ and S$_2$ states are almost degenerate, see  \reffig{C60-bond} c). Very rapidly in the simulation, the LUMO and LUMO+1 energy states show a strong mixing and a large NACV, and an electronic hopping between LUMO+1 and LUMO occurs (at $t = 4.75$ fs). The system evolves now in the $S_1$ excited state and the LUMO and LUMO+1 energy levels move away from each other. The LUMO orbital (occupied by one electron) has now the stronger bonding character, see  \reffig{C60-bond} a) ($t = 23$ fs). 
As shown in \reffig{C60-bond} d), in this part of the trajectory
there is a repulsive force $ F_{reac}$ between the two $C_{60}$ molecules but the distance $d_{reac}$ decreases. A detailed analysis of the motion of the system shows that while most of the carbon atoms in each molecule move away from the other molecule, the two frontiers atoms on each molecule actually approach closer towards the two frontier atoms on the other molecule due to the excited electron in the bonding LUMO orbital.  Consequently, both $C_{60}$ molecules tend to elongate. 
The repulsive force  $ F_{reac}$ and the distance $d_{reac}$ diminish in time and the system evolves towards a conical intersection between S$_0$ and S$_1$, which increases the NACV between the HOMO and LUMO. The proximity of the conical intersection induces an~electron switching between the LUMO and HOMO.
After this transition the HOMO and LUMO exchange their bonding and anti-bonding characters.  Now, the bonding HOMO has two electrons and the anti-bonding LUMO is empty. Consequently, strong chemical bonds are established between the frontier carbon atoms and the PCA reaction is completed. In the remaining simulation time the distance $d_{reac}$ oscillates around its equilibrium position while all the carbon atoms relax in a complex way to minimize the initial elongation.
   
An example of the non-reactive pathway is presented in \reffig{C60-dis}. Initially, the system evolves very similarly to the previous case where the system appraoches a conical intersection between S$_1$ and S$_2$;
due to the probabilistic nature of the fewest switches method, the electronic transition from LUMO+1 to LUMO takes place now a little later, at $t = 5.75$ fs, see \reffig{C60-dis} b). 
The system evolves then in the S$_1$ state, moving towards a S$_0$-S$_1$ conical intersection, but the electron hopping does not occur now. Thus, the systems stays in the S$_1$ PES, with a repulsive force $ F_{reac}$ acting between both molecules; the distance $d_{reac}$ oscillates (\reffig{C60-dis} d), due to the attraction associated with the bonding LUMO (occupied with one electron), see \reffig{C60-dis} a). During the second oscillation cycle, when $d_{reac}$
decreases again the S$_1$ and S$_0$ PESs (and the LUMO and HOMO energy levels) approach each other again and the electron hop takes place now, at time $t = 100$ fs (\reffig{C60-dis} b,c). The HOMO (occupied by two electrons) presents an anti-bonding character now (see \reffig{C60-dis} a), $t = 110$ fs), 
enhancing the repulsive force $ F_{reac}$. Consequently the distance $d_{reac}$ increases and the system goes to the weakly interacting ground state (\reffig{C60-PES}) and the PCA reaction does not occur.
    
\section{Conclusions}
The computational simulation of photo-induced processes in large molecular systems is a very challenging problem. Firstly, to properly simulate photo-induced reactions the PESs corresponding to excited states must be appropriately accessed; secondly, understanding the mechanisms of these processes requires the exploration of complex configurational spaces and the localization of conical intersections; finally, photo-induced reactions are probability events, that require the simulation of hundreds of trajectories to obtain the statistical information for the analysis of the reaction profiles.

Here, we have presented a detailed description of our implementation of a DFT-MDET technique within our local-orbital code {\sc Fireball}, suitable for the computational study of these problems. The MDET technique consists of two nested time loops: an outer loop for the motion the atoms and an inner loop for the propagation of the electronic states. The atoms move in adiabatic PESs, that are calculated using constrained DFT calculations for the different electronic configurations. The time evolution of the electronic states is calculated using time-dependent KS theory. The coupling of the atomic motion and the time evolution of the electronic states is taken into account by means of probabilistic hops between different PESs, using the fewest switches algorithm. When hops take place, energy conservation is imposed re-scaling the velocities along the direction of the non-adiabatic coupling vectors, that are calculated on the fly along the molecular dynamics simulation \cite{
Abad2013}.

As an example of the application of this DFT-MDET approach, we have presented some simulations for two different photo-induced cycloaddition reactions: the reaction of two small organic molecules (maleic anhydride and ethylene, forming cyclobutane-1-2-dicarboxilic anhydride) and the polymerization reaction of two C$_{60}$ molecules. In the first case, an electron is initially promoted from the HOMO to the LUMO; the LUMO presents a bonding character, inducing an attractive force between both molecules. 
At a time of $\sim$ 60 fs, the system goes through a S$_1$-S$_0$ conical intersection, with a large value of the NACV between HOMO and LUMO.  
We identify three different deactivation channels of the initial electron excitation, depending on the time of the electronic transition from LUMO to HOMO, and the character of the HOMO after the transition.
In particular, the cycloaddition reaction is established when the HOMO is a bonding state after the transition.
In the case of the larger C$_{60}$ molecules, the S$_1$ and S$_2$ states are initially almost degenerate and we promote one electron from the HOMO to the LUMO+1. Very rapidly in the simulations ($\sim$ 5 fs) the electron hops from the LUMO+1 to the LUMO and the system evolves in the S$_1$ state towards a S$_1$-S$_0$ conical intersection. In this part of the simulations there is an overall repulsive force between both molecules and an attractive force between the frontier atoms on each molecule due to the excited electron in the bonding LUMO state; thus, the molecules tend to elongate. We identify two different scenarios, depending on the time of the electronic transition from LUMO to HOMO and the character of the HOMO state after the transition: (i) a PCA reaction forming a bonded C$_{60}$ dimer; and (ii) the motion of the system towards the weakly coupled ground state.

\subsection*{Acknowledgments}
This work is supported by the Spanish MICIIN (project FIS2010-16046)

V.Z., P.H. and P.J. acknowledge the support by GA\v{C}R, grant no.\ 14-02079S.

\bibliographystyle{ieeetr} 
\bibliography{MDET}

\end{document}